\documentstyle[epsf,eqalign]{article}
\def\DM{\mbox{DM}}
\begin{document}
\title{The radioastronomical "Time Machine" effect and the
solution of gamma ray bursts mystery}
\author{ Galina V. Lipunova, Vladimir M. Lipunov,
       Ivan E. Panchenko\thanks{ Department of Physics, Moscow State University; 
Sternberg Astronomical Institute}}
\date{}
\maketitle

\markboth{G. Lipunova et al., : "Time Machine" solution of GRB}{}

\begin{abstract}

The possible low-frequency
radio emission from the progenitors of gamma ray bursts
can experience a delay from tens of seconds to hours
on the way to the observer due to the
dispersion in galactic and extragalactic plasma, and thus
reach the observer as a radio afterglow of the burst.
This opens a unique possibility (peculiar "Time Machine") of
seeing what happened in that
place before the catastrophe.

\end{abstract}

\section{Introduction}

It is well known, that if 
the gamma ray bursts (GRB) origin from the mergers of relativistic 
binary stars, their progenitors should be a strong source of gravitational
waves, observable by the detectors of the advanced LIGO type.

In addition, there are some reasons to expect that the progenitors of gamma-ray
bursts (GRB) could be also a 
powerful sources of radio emission. For example,
if GRB origin from binary neutron stars or neutron stars and black
holes mergers, their pre-merger evolution could be accompanied with
pulsar-like phenomena arising due to fast orbital motion of two compact
objects
(Lipunov \& Panchenko, 1996) or their magnitospheric interaction
(Vietri, 1996; Nulsen \& Fabian, 1984).  This emission could be observable
as a radio ``before-glow'' of GRB.

The observational detection of such radio sources is complicated
because their position on the sky is apriori unknown. But the time
delay that occurs in the propagation of low-frequency signal through
the galactic and extragalactic plasma makes it possible to observe the signal
emitted by the GRB progenitors {\it before} the burst {\it after } the
burst itself has been already observed and the position is known, thus
changing the beforeglow into afterglow.

\section{The radio signal delay}

The delay of a radio signal of a frequency $\nu$
passing a distance $dr$ through plasma with electron density $n_e$
equals to:
\begin{equation}
d t'_{\nu}= \frac{1}{2c}\frac{\nu_0^2}{\nu^2} dr,
\end{equation}
where $\nu_0$ is the plasma frequency
\begin{equation}
 \nu_0^2 = \frac{e^2 n_e(z)}{\pi m_e}.
\end{equation}

In the observer frame the same delay is multiplied by $(1+z)$.

Using  for estimate the simple case of a flat universe ($\Omega=1$)
with $\Omega_\Lambda = 0$,
in which the metric distance is
\begin{equation}
D_m(z)={2c\over H_0} \left(1-{1\over{\sqrt{1+z}}}\right)\,
\end{equation}
taking into account that the emission frequency changes while it
propagates through the Universe ($\nu(z)=\nu_{obs}(1+z)$)
and assuming that the relative density of extragalactic plasma $\Omega_p$
is constant within some $z$ which implies the electron density
\begin{equation}
  n_e(z)=n_e(0) (1+z)^3\
\end{equation}
at a redshift $z$,
we obtain the delay
\begin{equation}
   \Delta t_{\nu}=
   \frac{1}{2c} \frac{e^2}{\pi m_e \nu_{obs}^2 }\DM
    \approx 0.42\nu_8^{-2} \DM, 
\end{equation}
where $\nu_8=\nu_{obs}/100$~MHz and the dispersion measure is
\begin{equation}
  \eqalignleft{
   \DM &=  \int\limits_0^z \frac{\ n_e(z) dD_m}{\ 1+z} \cr 
    &  = \frac{\ H_0c\Omega_m}{\ 4\pi G m_p} \left( (1+z)^{3/2} -1 \right)\cr 
  }
\end{equation}
where
$\Omega_{0.01}=\Omega_p/0.01$ and $h_{50}=H_0/75$~km/s/Mpc.

Now the delay can be rewritten as
\begin{equation}
   \eqalignleft { 
    \Delta t_\nu & =
   \frac{\displaystyle e^2 H_0 \Omega_m}{\displaystyle 8\pi G m_p m_e \nu_{obs}^2}
         \left( (1+z)^{3/2} -1 \right) \cr
    &
   \approx 75 \Omega_{0.01} h_{75}\nu_8^{-2} \left( (1+z)^{3/2} -1 \right)\mbox{s} .
 } 
\end{equation}

Of course, the delay takes place not only in the extragalactic media,
where the electron density is
$6.5\cdot 10^{-8} h_{75}^2\Omega_{0.01}$,
 but
also in our Galaxy, and in the host galaxy.
For example, our galaxy provides the dispersion measures from $16.5$ in
the polar direction to $1400$ in the direction of its center
(Taylor, Cordes, 1993).
The contribution of the host galaxy can be in the same range, or
even higher because of its lower age.

Thus, we can expect that at a frequency of
$100$~MHz the delay will be of the order of minutes or tens of minutes,
and even hours at $10$~Mhz.
Therefore, the gamma-ray burst allows us to observe up to an hour of
its radio prehistory.

\section{The possible radio flux}

Let us estimate the radio flux on the base of magnetic dipole energy losses.

The spectral flux density observered at the frequency $\nu_{obs}$ if
the source has a power-law spectrum with the index $n$
and is located at a redshift $z$, is equal to

\begin{equation}
F_{\nu_{obs}} = \frac{L_{\nu_{obs}}}{4\pi D_m(z)^2 (1+z)^{2+n}},
\end{equation}
where $L_{\nu_{obs}}$ is the spectral luminosity of the source
at the frequency $\nu_{obs}$.

For the purpose of estimating the flux, let us assume that
$L_\nu\sim \eta L/\nu$, where $L$ is the total energy output
of the source. Then $\eta$ has the meaning of the "efficiency of
radioemission". If we approximate the total luminosity by the
magnetic dipole formula, then we would obtain the flux
\begin{equation}
\eqalignleft { 
F_{\nu_{obs}} &= \frac {\mu^2\pi^3\nu_{spin}^4\eta H_0^2 }{c^5 \nu_{obs}} K(z) =\cr
              &=   3.7 \mbox{Jy} \frac {\eta_{-4} h_{50}^2}{P_{-3}^4\nu_{obs}} K(z),
}
\end{equation}
where $ \eta_{-4}=\eta/10^{-4}$, $P_{-3}$ is the dipole spin period
in milliseconds, and
\begin{equation}
     K(z)=\left((1+z)\left(1-\frac{1}{\sqrt{1+z}}\right)\right)^{-1} .
\end{equation}

This flux estimate is rather high, which means that even for less
efficient energy losses than a dipole one the value of the flux can be
over the limit of detection.

\section{Discussion}
 As shown in previous sections, the dispersion of the low-frequency emission
during its propagation in plasma can lead to the "time-machine" effect,
allowing us to observe several minutes of the
radio prehistory of a GRB at $100$~Mhz, or even hours at $10$~Mhz.
As shown by Lipunov \& Panchenko (1996), this last minute can be
accompanied by powerful pulsar-like emission, arising from the fast
orbital motion of two magnetic dipoles with millisecond periods.
According to the well-known formula, describing the gravitational wave
lifetime of a binary (Shapiro \& Teukolsky, 1983):
\begin{equation}
T_0 = 10^5 P_{orb}^{8/3} \mbox{s}, \label{T(p)}
\end{equation}
a minute before the merger (which is associated with the GRB itself)
the orbital period is $\approx 10$~ms, and a hour before it is
$\approx 140$~ms.
The pre-GRB emission should (see  Lipunov \& Panchenko (1996), Fig.~2)
display a quick burst-like power-law growth of intensivity,  
and should probably be modulated with the decreasing as given by
(\ref{T(p)}) period.  

In general, this emission could be observed as
a ``beforeglow'' of a GRB together with its gravitational wave ``beforeglow'', 
but if the dispersion measure is high enough,
the time delay may turn it into its afterglow.  
 
Due to the fact that this effect has a strong observational frequency
dependance ($\sim \nu_{obs}^{-2}$), it could be preferably 
detected by the large phased arrays 
similar to those of 
Pushino and Kharkov radio astronomy observatories 
Kharkov ,
which are sensitive in the frequency range $10-100$~Mhz.

The importance of the described effect of "radio-astronomical time machine"
is that it provides a new tool for studying GRB by opening a direct view 
of their precursors.
Also search for it can involve new kinds of radio instruments in the GRB
follow-up observations programme.

\vspace{2cm}
{\it
Authors acknowledge their colleagues Drs K.A. Postnov and M.E. Prokhorov for
useful discussions.
}

\end{document}